\documentstyle{article}

\title{Quantum Mechanics as a Classical Theory I:\\
Non-relativistic Theory}

\author{L.S.F. Olavo\\
Departamento de Fisica, Universidade de Brasilia,\\
70910-900 - Brasilia - D.F. - Brazil}

\date{march, 30, 1995}

\begin{document}

\maketitle

\begin{abstract}
The objective of this series of three papers is to axiomatically derive
quantum mechanics from classical mechanics and two other basic axioms. In
this first paper, Schroendiger's equation for the density matrix is fist
obtained and from it Schroedinger's equation for the wave functions is
derived. The momentum and position operators acting upon the density matrix
are defined and it is then demonstrated that they commute. Pauli's equation
for the density matrix is also obtained. A statistical potential formally
identical to the quantum potential of Bohm's hidden variable theory is
introduced, and this quantum potential is reinterpreted through the
formalism here proposed. It is shown that, for dispersion free {\it ensembles%
}, Schroedinger's equation for the density matrix is equivalent to Newton's
equations. A general non-ambiguous procedure for the construction of
operators which act upon the density matrix is presented. It is also shown
how these operators can be reduced to those which act upon the wave
functions.
\end{abstract}

\section{General Introduction}

Ever since the establishment of the Copenhagen Interpretation and it's
purely epistemological point of view, quantum mechanics has been subject to
criticism, debates and of constant revisionary attempts trying to alter it
in one way or another\cite{1}-\cite{7}. Other than the hidden variables
theories,
few attempts were made to reconstruct quantum theory on the lines of
Realism. Even these theories, according to Bell's theorem\cite{8}-\cite{10},
must posses non-local characteristics little acceptable to a realist such as
Einstein\cite{1}.

Part of the problem resides in the fact that, unlike relativistic theories,
it has been hitherto impossible to derive quantum mechanics from classical
mechanics though the insertion of a few complementary postulates capable of
identifying the differences between both theories. The absence of this
derivation has led most physicists to believe in the existence of an
unbridgeable conceptual abyss separating the two theories\cite{11}. Because
of this, most of the ontologies proposed for quantum mechanics reject
Realism {\it a priori}.

In this series of papers we intend to demonstrate that both relativistic and
non-relativistic quantum mechanics can be derived from classical mechanics
thorough the addition of two somewhat ''natural'' postulates which do not
alter the classical character of the derivation. We intend to demonstrate
that, based on a rather simple generalization of these postulates, it is
possible to construct a relativistic general quantum theory for {\it %
ensembles} of single particle systems.

One of the most important results, at least where epistemology is concerned,
is the mathematically backed negation of the ontological origin of
Heisemberg's uncertainty relations. This is the key to refuting the
Copenhagen Interpretation as a whole together with it's ontology\cite{12}.
Also in the relation to epistemology, we propose a quantum model in which
the observer is included without its {\it ad hoc} postulation being necessary%
\cite{13}. From there, we maintain that a general measurement theory in
quantum mechanics (and in all of physics) is impossible, analyzing von
Neumann's\cite{14} attempt as an example.

In the same manner, and independently of Bell's argument\cite{9,15},
which we reject, we demonstrate the errors in von Neumann's theorem on the
impossibility of hidden variable theories. Our counter example is a theory
of local behavior. We show that Bell's theorem does not necessarily entail
in non-local quantum mechanics and that quantum mechanics' hidden variable
theory is, in fact, newtonian mechanics itself. We also take this
opportunity to demonstrate that Bohm's\cite{7,16,17} theory is not truly a
hidden variable theory and reinterpret its formalism.

In a totally formal perspective, we show that Schroedinger's equation for
the density matrix is fundamental, and not that for the probability
amplitudes. Also, Dirac's second order equation is shown to be the
fundamental one (and not his linear equation)\cite{18}. We also present a
general technique for the construction of operators\cite{19}.

We therefore demonstrate that, starting from our basic axioms, it is
possible to obtain all of quantum mechanics and much more, and that this
theory is nothing more than a classical theory.

This series of papers is divided in the following manner:

In this first paper we develop non-relativistic formalism in it's most
fundamental results, obtaining Schroedinger's equations for the density
matrix (which we call the density function) and for the wave function (which
we call the probability amplitude). We develop the concept of operators
which act upon the density function and upon the probability amplitude, and
demonstrate that the first commute according to different rules from the
latter - in the appendix we present a general technique free from
ambiguities for obtaining operators. Pauli's equation is also obtained by
including internal degrees of freedom associated to the intrinsic magnetic
moment. The basic equations for Bohm's hidden variable theory are also
found, but only shortly discussed. We also demonstrate how to introduce the
observer into quantum formalism. The epistemological implications of our
results are left for the third paper of this series.

In the second paper we derive Schroedinger's relativistic equations for the
density function and for the probability amplitude through small
modifications in the axioms of the first paper, making them coherent with
the special theory of relativity. The calculations of the first paper are
repeated in order to find a relativistic theory for one particle. We also
find a system of equations involving Einstein's which take into account the
effects of gravitation.

The third and final paper of this series discusses the epistemological
implications of the results of the first two papers. Based upon the
clarification of Heisemberg's misconception interpretation of the
uncertainty relations, we construct a Realistic Interpretation to contrast
with Copenhagen's. Von Neumann's measurement theory is discussed in
comparison to the results of our first paper through Realistic epistemology.
We demonstrate that von Neumann's theorem on the impossibility of a hidden
variable theory is not correct but that, on the other hand, Bell's
argumentation is also unacceptable. To be more precise, we demonstrate that,
taking the non-relativistic case as an example, newtonian theory is the
theory of hidden variables of quantum mechanics. Bell's theorem is analyzed
and we demonstrate that it does not imply in a non-locality of quantum
mechanics. We discuss Bohm's theory and demonstrate that it cannot be a
hidden variable theory, since it is not free of the dispersions associated
with Heisemberg's relations.

\section{Introduction}

In this paper, we obtain non-relativistic quantum theory's mathematical
formalism from newtonian mechanics and two additional postulates. For
reasons of clarity, these results are presented in an axiomatic way.

In the third section, we present the basic formalism, deriving
Schroedinger's equations for the density function and for the probability
amplitude. We introduce the operator concept for these two functions and
demonstrate that they obey distinct commutation laws. From there we show
that Heisemberg's uncertainty relations are not ontological, but purely
formal.

In the fourth section, Pauli's equation for the density function is derived
taking into consideration the particle's internal magnetic moments.

In the fifth section, we introduce and interpret the idea of statistical
potential, formally similar to Bohm's quantum potential.

The sixth section proposes a quantum experiment arrangement where a physical
system acting as an external observer is introduced and its influence can be
formally taken into account.

In the last section, we present our conclusions.

The appendix A discusses some of Wigner-Moyal Infinitesimal Transformation's
properties of the state functions $F\left( x,p;t\right) $. It is also
demonstrated that Schroedinger's equation for the density function can be
reduced to Newton's equations in the limit of dispersion free {\it ensembles}%
{}.

Appendix B brings a general free from ambiguities method for the
construction of operators.

In appendix C the calculations made in the body of the work are generalized
for three-dimensional systems composed of many particles.

Appendix D is concerned with the definition of {\it mixed states} and the
introduction of the concept of {\it density matrix}. The relation between
the calculations made in the main text and the trace operation upon the
density matrix is also investigated.

In both this paper and the other two we will use the terms ''classical'' and
''quantum'' meaning a classical statistical mechanics built upon phase or
configuration space, respectively (as will become clear ahead).

\section{Axioms and Formalism}

We begin our theory with {\it ensembles} described by probability density
functions in classical phase space written as $F\left( x,p;t\right) $. The
variables which label this function represent the position and the momentum
of the particles that make up the {\it ensemble}.

Let us now list the theory's axioms:

\begin{description}
\item[(A1)]  Newtonian particle mechanics is valid for all particles which
constitute the systems which compose the {\it ensemble}.

\item[(A2)]  For an isolated system the joint probability density function
is conserved:
\begin{equation}
\label{(1)}\frac{dF\left( x,p;t\right) }{dt}=0
\end{equation}

\item[(A3)]  The Wigner-Moyal Infinitesimal Transformation defined as

\begin{equation}
\label{(2)}\rho \left( x+\frac{\delta x}2,x-\frac{\delta x}2;t\right) =\int
F\left( x,p;t\right) \exp \left( i\frac{p\delta x}\ell \right) dp
\end{equation}
where $\ell $ is a universal parameter (having the same value for all
transformations) with dimensions of angular momentum, is adequate for the
description of any non-relativistic quantum system (this definition differs
from that usually made in the literature\cite{20}-\cite{22} in its
infinitesimal nature just emphasized).
\end{description}

We now demonstrate that all non-relativistic quantum mechanics, and some
additional results, can be obtained from these axioms.

Using equation (\ref{(1)}) we have
\begin{equation}
\label{(3)}\frac{dF\left( x,p;t\right) }{dt}=\frac{\partial F}{\partial t}+
\frac{dx}{dt}\frac{\partial F}{\partial x}+\frac{dp}{dt}\frac{\partial F}{%
\partial p}=0.
\end{equation}

We can use Newton's equations, axiom (A1), on (\ref{(3)})
\begin{equation}
\label{(4)}\frac{dx}{dt}=\frac pm\quad ;\quad \frac{dp}{dt}=f,
\end{equation}
supposing that the force $f$ derives from a potential $V\left( x\right) $:
\begin{equation}
\label{(5)}f=-\frac{\partial V}{\partial x}.
\end{equation}
Multiplying this equation by the exponential in (\ref{(2)}) and integrating,
we reach
\begin{equation}
\label{(6)}-\frac{\partial \rho }{\partial t}+\frac{i\ell }m\frac{\partial
^2\rho }{\partial x\partial \left( \delta x\right) }-\frac i\ell \delta
V\rho =0,
\end{equation}
where we use the infinitesimal character of $\delta x$ to write
\begin{equation}
\label{(7)}\frac{\partial V}{\partial x}\delta x=\delta V\left( x\right)
=V\left( x+\frac{\delta x}2\right) -V\left( x-\frac{\delta x}2\right)
\end{equation}
and also from the fact that
\begin{equation}
\label{(8)}\left[ F\left( x,p;t\right) \exp \left( \frac{ip\delta x}\ell
\right) \right] _{p=-\infty }^{p=+\infty }=0,
\end{equation}
as is expected from a probability density function

Changing the variables
\begin{equation}
\label{(9)}y=x+\frac{\delta x}2\quad ;\quad y^{\prime }=x-\frac{\delta x}2
\end{equation}
we can rewrite equation (\ref{(6)}) above as
\begin{equation}
\label{(10)}\left\{ \frac{\ell ^2}{2m}\left[ \frac{\partial ^2}{\partial y^2}%
-\frac{\partial ^2}{\partial y^{\prime 2}}\right] -\left[ V\left( y\right)
-V\left( y^{\prime }\right) \right] \right\} \rho \left( y,y^{\prime
};t\right) =-i\hbar \frac \partial {\partial t}\rho \left( y,y^{\prime
};t\right) ,
\end{equation}
which we call Schroedinger's First Equation for the density function $\rho
\left( y,y^{\prime };t\right) $. This equation is valid for all values of $y$
and $y^{\prime }$, as long as these are infinitesimally close, as can be
seen in (\ref{(9)}). It must not be forgotten that this consideration is not
inconvenient to our calculations, for reasons that will become clear ahead.
We have proven the following theorem:

\begin{description}
\item[(T1)]  The density function $\rho \left( y,y^{\prime };t\right) $
satisfies Schroedinger's First Equation (\ref{(10)}) being $\ell $ a
parameter to be determined experimentally (we know that this parameter is
Plank's constant and thus we will henceforth write $\hbar $ instead of $\ell
$).
\end{description}

Let us now suppose that we can write
\begin{equation}
\label{(11)}\rho \left( y,y^{\prime };t\right) =\Psi ^{*}\left( y^{\prime
};t\right) \Psi \left( y;t\right) ,
\end{equation}
where we will call $\Psi \left( y;t\right) $ the probability amplitude.
Since this is usually a complex function, we can write it as
\begin{equation}
\label{(12)}\Psi \left( y;t\right) =R\left( y;t\right) \exp \left[ iS\left(
y;t\right) /\hbar \right] ,
\end{equation}
where $R\left( y;t\right) $ and $S\left( y;t\right) $ are real functions.

We can now expand the function $\rho \left( y,y^{\prime };t\right) $, given
in equation (\ref{(11)}) in terms of $x$ and $\delta x$, to obtain%
$$
\rho \left( x+\frac{\delta x}2,x-\frac{\delta x}2;t\right) =
$$
\begin{equation}
\label{(13)}=\left\{ R\left( x;t\right) ^2-\left( \frac{\delta x}2\right)
^2\left[ \left( \frac{\partial R}{\partial x}\right) ^2-R\left( x;t\right)
\frac{\partial ^2R}{\partial x^2}\right] \right\} \exp \left( \frac i\hbar
\delta x\frac{\partial S}{\partial x}\right)
\end{equation}
which, substituted in equation (\ref{(6)}), yields%
$$
\left[ \frac{\partial \left( R^2\right) }{\partial t}+\frac \partial
{\partial x}\left( R^2\frac{\partial S/\partial x}m\right) \right] +
$$
\begin{equation}
\label{(14)}+\frac{i\delta x}\hbar \left\{ \frac{\hbar ^2}{2mR^2}\frac
\partial {\partial x}\left[ \left( \frac{\partial R}{\partial x}\right) ^2-R
\frac{\partial ^2R}{\partial x^2}\right] +\frac \partial {\partial x}\left[
\frac 1{2m}\left( \frac{\partial S}{\partial x}\right) ^2+V+\frac{\partial S
}{\partial t}\right] \right\} =0.
\end{equation}
Canceling the real and complex terms, we obtain the following equations
\begin{equation}
\label{(15)}\frac{\partial P}{\partial t}+\frac \partial {\partial x}\left(
P \frac{\partial S/\partial x}m\right) =0,
\end{equation}
\begin{equation}
\label{(16)}\frac \partial {\partial x}\left[ \frac 1{2m}\left( \frac{%
\partial S}{\partial x}\right) ^2+V+\frac{\partial S}{\partial t}-\frac{%
\hbar ^2}{2mR}\frac{\partial ^2R}{\partial x^2}\right] =0,
\end{equation}
where
\begin{equation}
\label{(17)}P\left( x;t\right) =R\left( x;t\right) ^2=\lim _{\delta
x\rightarrow 0}\rho \left( x+\frac{\delta x}2,x-\frac{\delta x}2;t\right)
\end{equation}
is the probability density in configuration space, as becomes clear if we
look at expression (\ref{(2)}). This last equation justifies the comment
made after (\ref{(10)}); the mean values of the quantities will always be
calculated within the limit given in (\ref{(17)}) so that the calculation of
the density function for infinitesimally close points can be done without
the loss of generality. (for mixtures, see appendix D). This does not imply
that only the element for which $\delta x$ is equal to zero contributes. The
kinematical evolution of the density function is governed by equation (\ref
{(6)}) which mixes all the contributions. The above mentioned limit must be
taken after this equation has been solved.

Equation (\ref{(16)}) can be rewritten as
\begin{equation}
\label{(18)}\frac 1{2m}\left( \frac{\partial S}{\partial x}\right) ^2+V+
\frac{\partial S}{\partial t}-\frac{\hbar ^2}{2mR}\frac{\partial ^2R}{%
\partial x^2}=const.
\end{equation}
We can obtain the constant's value considering the solution for a free
particle. Calculations done, it is easy to demonstrate that this constant
should cancel itself. With this constant equal to zero, equation (\ref{(16)}%
) is equivalent to
\begin{equation}
\label{(19)}\frac{\hbar ^2}{2m}\frac{\partial ^2\Psi }{\partial x^2}-V\left(
x\right) \Psi =-i\hbar \frac{\partial \Psi }{\partial t}
\end{equation}
since, if we substitute the decomposition made in expression (\ref{(12)}) in
the equation above, we obtain (\ref{(16)}) once again. We will call equation
(\ref{(19)}) Schroedinger's Second Equation for the probability amplitude.
It is thus demonstrated that

\begin{description}
\item[(T2)]  If we can write the density function $\rho $ as the product (%
\ref{(11)}), than the probability amplitude $\Psi \left( x;t\right) $
satisfies Schroedinger's Second Equation (\ref{(19)}) together with the
Equation of Continuity (\ref{(15)}). (We will justify the nomenclature given
to equation (\ref{(15)}) further on).
\end{description}

We introduce the operator concept through the formal identification (an
apostrophe will always be put in order to distinguish operators which act
upon the density function from those that act upon the probability
amplitude)
\begin{equation}
\label{(20)}\stackrel{\wedge }{p}^{\prime }=-i\hbar \frac \partial {\partial
\left( \delta x\right) }\quad ;\quad \stackrel{\wedge }{x}^{\prime }=x,
\end{equation}
based on the fact that
\begin{equation}
\label{(21)}\overline{p}=\lim _{\delta x\rightarrow 0}-i\hbar \frac \partial
{\partial \left( \delta x\right) }\int F\left( x,p;t\right) \exp \left( i
\frac{p\delta x}\hbar \right) dxdp
\end{equation}
and
\begin{equation}
\label{(22)}\overline{x}=\lim _{\delta x\rightarrow 0}\int xF\left(
x,p;t\right) \exp \left( i\frac{p\delta x}\hbar \right) dxdp.
\end{equation}

Thus,

\begin{description}
\item[(T3)]  The result of the operation upon the density function $\rho $
of the momentum and position operators, defined by the expressions in (\ref
{(20)}), represents, respectively, the mean values for position and momentum
for the {\it ensembles'} components.
\end{description}

Remember that, using the expansion of the density function given in (\ref
{(13)}), we have
\begin{equation}
\label{(23)}\overline{p}=\lim _{\delta x\rightarrow 0}-i\hbar \frac \partial
{\partial \left( \delta x\right) }\int F\left( x,p;t\right) \exp \left( i
\frac{p\delta x}\hbar \right) dxdp=\int R\left( x\right) ^2\left( \frac{%
\partial S}{\partial x}\right) dx,
\end{equation}
and this justifies our calling equation (\ref{(15)}) the Continuity Equation.

In order to obtain the momentum operator action upon the probability
amplitude we can rewrite the equation above as
\begin{equation}
\label{(24)}\lim _{\delta x\rightarrow 0}-i\hbar \int \frac \partial
{\partial \left( \delta x\right) }\left[ \Psi ^{*}\left( x-\frac{\delta x}%
2;t\right) \Psi \left( x+\frac{\delta x}2;t\right) \right] dx
\end{equation}
and reach, after some calculations, the result
\begin{equation}
\label{(25)}Re\left\{ \int \Psi ^{*}\left( x;t\right) \left( -i\hbar \frac
\partial {\partial x}\right) \Psi \left( x;t\right) dx\right\} .
\end{equation}
The same can be done for the position operator (it is worth stressing that
the hermitian character is automatically established). This allows us to
define the position and momentum operators
\begin{equation}
\label{(26)}\stackrel{\wedge }{p}\Psi \left( x;t\right) =-i\hbar \frac
\partial {\partial x}\Psi \left( x;t\right) \quad ;\quad \stackrel{\wedge }{x%
}\Psi \left( x;t\right) =x\Psi \left( x;t\right)
\end{equation}
as usual. Defining the Hamiltonian operator as
\begin{equation}
\label{(27)}\stackrel{\wedge }{H}\Psi \left( x;t\right) =\left[ \frac{p^2}{2m%
}+V\left( x\right) \right] \Psi \left( x;t\right) =i\hbar \frac \partial
{\partial t}\Psi \left( x;t\right) ,
\end{equation}
we can rewrite Schroedinger's Second Equation in operator terms such as
\begin{equation}
\label{(28)}\stackrel{\wedge }{H}\Psi \left( x;t\right) =i\hbar \frac
\partial {\partial t}\Psi \left( x;t\right) .
\end{equation}

If we now define the commutator of two operators in the usual form, it
becomes clear that
\begin{equation}
\label{(29)}\left[ \stackrel{\wedge }{x},\stackrel{\wedge }{p}\right]
=i\hbar .
\end{equation}

{}From this result, it is easy to show that the relation
\begin{equation}
\label{(30)}\Delta x\Delta p\geq \hbar /2
\end{equation}
known as Heisemberg's uncertainty relation, must be valid. Before this
result is interpreted, it is necessary to stress that
\begin{equation}
\label{(31)}\left[ \stackrel{\wedge }{x}^{\prime },\stackrel{\wedge }{p}%
^{\prime }\right] =0
\end{equation}
and, therefore, that we will have the following relation associated with
these operators:
\begin{equation}
\label{(32)}\Delta x\Delta p\geq 0.
\end{equation}

This result was, in fact, expected since no hypothesis about the mean
squared deviations associated to the (classical) function $F\left(
x,p;t\right) $ were made. It demonstrates that the relation (\ref{(30)})
results from the recognition of the possibility of writing the density
function $\rho $ as the product represented in (\ref{(11)}). In this manner,
far from representing a fundamental property of nature, relation (\ref{(30)}%
) represents a limitation of our descriptions according to equation (\ref
{(19)}). We put this as a theorem:

\begin{description}
\item[(T4)]  Quantum mechanics, as developed according to equation (\ref
{(19)}), is only applicable to problems where the density function $\rho $
can be decomposed according to (\ref{(11)}). In these cases, the product of
the mean quadratic deviations in the position and momentum of a system
represented by the joint probability density function $F\left( x,p;t\right) $
is such that

$$
\Delta x\Delta p\geq \hbar /2
$$
and has a lower limit.
\end{description}

We here note that manuals commonly assume that the Second Schroedinger's
Equation for the probability amplitude is the fundamental one to be obtained
and that, writing the density function as in (\ref{(11)}), the First
Schroedinger's Equation is derived. We demonstrate here that this sequence
is unjustifiable. While the equation for the amplitudes presents us with a
dispersion relation such as in (\ref{(30)}), the equation for the density
function is dispersion free (see appendix D).

The results obtained above are the foundation of the whole non-relativistic
quantum mechanical formalism, which we will not derive again.

Let us pass on to the derivation of Pauli's equation

\section{Pauli's Equation}

Up to this point we have discussed only systems constituted of particles
with no internal degree of freedom. Consider now the case in which the
system's components possess an intrinsic magnetic moment capable of coupling
to an external magnetic field. We can expand this field around the region
occupied by the particle
\begin{equation}
\label{(33)}{\bf H}\left( {\bf x}\right) ={\bf H}\left( {\bf x}_0\right)
+\left( {\bf x}-{\bf x}_0\right) \cdot \nabla {\bf H}\left( {\bf x}_0\right)
+...,
\end{equation}
where $x_0$ is the particle's position. The coupling between the particle's
intrinsic magnetic moment and the magnetic field is
\begin{equation}
\label{(34)}{\bf F}_m=-\nabla \left( {\bf m}\cdot {\bf H}\right) +..\quad .
\end{equation}

Newton's equation for this system can be written as
\begin{equation}
\label{(35)}\frac{d{\bf p}}{dt}={\bf f}_{mec}+{\bf F}_m,
\end{equation}
where ${\bf f}_{mec}$ represents general mechanical forces (derivable from a
potential $V\left( x\right) $). Following similar steps to those used in the
previous section and in appendix C, we reach the following equation for the
density function
\begin{equation}
\label{(36)}\left\{ \frac{\hbar ^2}{2m}\left[ \nabla _y^2-\nabla _{y^{\prime
}}^2\right] -\left[ V\left( y\right) -V\left( y^{\prime }\right) \right]
-\left[ {\bf m}\cdot {\bf H}\left( y\right) -{\bf m}\cdot {\bf H}\left(
y^{\prime }\right) \right] \right\} \rho =-i\hbar \frac \partial {\partial
t}\rho
\end{equation}
which we call Pauli's Equation for the density function\cite{23}.

The extra degrees of freedom are represented by quantities $m_i$, which we
do not know. To obtain information about these quantities, we can consider
the precession equation which they should obey
\begin{equation}
\label{(37)}\frac{dm_i}{dt}=\epsilon _{ijk}m_kH_j,
\end{equation}
where $\epsilon _{ijk}$ is the totally anti-symmetric tensor. Comparing
these equations with those that we obtained when we wrote $m_i$ as
operators, we obtain the following commutation relation:
\begin{equation}
\label{(38)}\frac 1{i\hbar }\left[ \stackrel{\wedge }{m}_i,\stackrel{\wedge
}{m}_j\right] =\epsilon _{ijk}\stackrel{\wedge }{m}_k
\end{equation}
from which we can construct an appropriate matrix representation.

We can still write
\begin{equation}
\label{(39)}\stackrel{\wedge }{\bf m}=g\frac e{2mc}\stackrel{\wedge }{\bf S}%
\quad ;\quad \stackrel{\wedge }{\bf S}=\frac \hbar 2\stackrel{\wedge }{%
\sigma },
\end{equation}
where {\it g} is the Land\'e factor. Obviously, we can not know the value of
this factor until we study the relativistic problem.

We have obtained the following result:

\begin{description}
\item[(T5)]  A body with internal magnetic moment capable of coupling to an
external magnetic field obeys Pauli's Equation (\ref{(36)}).
\end{description}

In the second paper of this series we approach the general and special
relativistic quantum mechanical problem of {\it ensembles} composed of
single particle systems. We will also derive Dirac's and Klein Gordon's
Equations in addition to a general relativistic quantum equation.

\section{The Statistical Potential}

We can return to equation (\ref{(18)})
\begin{equation}
\label{(40)}\frac 1{2m}\left( \frac{\partial S}{\partial x}\right) ^2+V+
\frac{\partial S}{\partial t}-\frac{\hbar ^2}{2mR}\frac{\partial ^2R}{%
\partial x^2}=0
\end{equation}
and note that it may be considered a Hamilton-Jacobi equation for one
particle subjected to an effective potential
\begin{equation}
\label{(41)}V_{eff}\left( x\right) =V\left( x\right) -\frac{\hbar ^2}{2mR}
\frac{\partial ^2R}{\partial x^2}.
\end{equation}

Thus, we can formally write
\begin{equation}
\label{(42)}\frac{dp}{dt}=-\frac \partial {\partial x}V_{eff}\left( x\right)
=-\frac \partial {\partial x}\left[ V\left( x\right) -\frac{\hbar ^2}{2mR}
\frac{\partial ^2R}{\partial x^2}\right] ,
\end{equation}
along with the initial condition
\begin{equation}
\label{(43)}p=\frac{\partial S}{\partial x}.
\end{equation}

The integration of the system (\ref{(42)}), (\ref{(43)}) will give us a
series of trajectories which will be equivalent to the ''force'' lines
associated to the effective potential (\ref{(41)}). The resolution method
for this system of equations is as follows: first Schroedinger's equation
must be solved in order to obtain the probability amplitudes referent to the
{\it ensemble}. Once these amplitudes have been obtained, the effective
potential, which will act as a {\it statistical field for the ensemble}, is
built. One rather instructive example of this calculation is it's
application to the double slit experiment\cite{24} (an example of a time
dependent problem can be found in the literature\cite{25}.

It must be stressed that the equations for the individual constituents of
the system are Newton's equations. Thus, the potential (\ref{(41)}) must not
be considered a real potential, but a fictitious potential which acts as a
field in reproducing, through ''trajectories'', the statistical results of
the original equation (\ref{(40)}). We see this potential as a {\it %
statistical potential}.

The discussion of Bohm's hidden variable theory\cite{7} along this
reinterpretation of the potential (\ref{(41)}) will be undertaken in the
last paper of this series.

\section{The Observer}

With the advent of quantum mechanics, the question of the observer began to
occupy a prominent position in physics. Treated with mathematical rigor for
the first time by von Neumann\cite{14}, and later by a series of
authors\cite{4,5,17,26,27}, the observer acquired elevated epistemological
{\it status} through quantum theory, even though his function within this
theory's formalism is rather disputable\cite{13}, since the variables to him
associated never occur within the formalism itself (this will be further
discussed in the third paper of this series).

The various measurement theories which have been proposed have profound
epistemological and philosophical implications. One example is the
Copenhagen Interpretation's inevitable conclusion that the observer's
consciousness is necessary in the measuring process - as it is responsible
for the collapse of the state vector when an observation is made\cite{14,27}.
We shall present a model in which the observer, seen as a physical system,
may be introduced into quantum theory. We are not suggesting a general
measurement theory, as did von Neumann, but analyzing a specific physical
system. Nevertheless, the model here proposed will give us conditions to
discuss the philosophical framework of measurement theories proposed for
quantum theory.

It must be noted that the fundamental equation used, Liouville's equation in
phase space, is valid only for a closed system. Such a system\cite{4} may
have its behavior described by Schroedinger's First equation, but then there
is no collapse of the state vector (no observer at all). In this manner, the
state vector's collapse must always be postulated as a consequence of the
intervention of a ghost external observer, as it is by von Neumann, for
example. And more, since this observer could be considered a member of a
larger system, we are forced, in traditional analysis, to an infinite
regression which stops only when an observer with a consciousness capable of
such a reduction is postulated\cite{27}.

Yet let us see how we can treat a particular case in the perspective of
Realism.

Consider a two slit experiment as shown in figure 1. In this experiment the
{\it source1} sends {\it type1} particles through one of the first sources
slits. These particles have distribution $F_1^0\left( {\bf x},{\bf p}%
;t\right) $ when they leave {\it source1} and, without the influence of
external factors, would have a $F_1\left( {\bf x},{\bf p};t\right) $
distribution measured by the {\it detectors1} of the second screen. But let
us say that we desire to measure this distribution in a position previous to
the second screen (a typically inter phenomenon measure). In this case we
position {\it source2}, capable of emitting {\it type2} particles, as is
shown in figure 1. These particles exit {\it source2} with the distribution $%
F_2^0\left( {\bf x},{\bf p};t\right) $ and if it was not for the {\it type1}
particles they would hit {\it detectors2} with the distribution $F_2\left(
{\bf x},{\bf p};t\right) $. According to the hypothesis of Realism, we can
expect that the alteration of these two distributions will be the product of
collisions between both system's particles (considered distinct here for
simplicity). So, after the collisions take place we expect to find {\it %
system1} and {\it system2} represented by the distributions $F_1^1\left(
{\bf x},{\bf p};t\right) $ and $F_2^1\left( {\bf x},{\bf p};t\right) $
respectively.

We expect, as we do for Boltzmann's Equation\cite{28}, that the equation
satisfied by the new $F_1^1$ distribution be now given as
\begin{equation}
\label{(44)}\frac{dF_1^1}{dt}=\frac{\partial F_1^1}{\partial t}+\frac{{\bf p}%
_1}{m_1}\cdot \nabla _xF_1^1+\frac{d{\bf p}_1}{dt}\cdot \nabla
_pF_1^1=D_cF_1,
\end{equation}
where $D_CF_1$ represents the change in distribution $F_1$ due to the
collisions. We can divide this change according to two sources: one caused
by the collisions which fling the particles into the phase space element,
which we call $D_C^{\left( +\right) }F_1d{\bf x}_1d{\bf p}_1$, and one that
flings them out of this volume, which we call $D_C^{\left( -\right) }F_1d%
{\bf x}_1d{\bf p}_1$. It is clear that
\begin{equation}
\label{(45)}D_CF_1=D_C^{\left( +\right) }F_1-D_C^{\left( -\right) }F_1.
\end{equation}

The term $D_C^{\left( -\right) }F_1d{\bf x}_1d{\bf p}_1$ can be calculated
once considered that, within the volume element, the probability of a
collision sending particles outside this volume is
\begin{equation}
\label{(46)}\sigma \left( {\bf p}_1,{\bf p}_2\rightarrow {\bf p}_1^{\prime },%
{\bf p}_2^{\prime }\right) d^3p_1^{\prime }d^3p_2^{\prime },
\end{equation}
where $\sigma \left( {\bf p}_1,{\bf p}_2\rightarrow {\bf p}_1^{\prime },{\bf %
p}_2^{\prime }\right) $ is the cross section for the collisions in which the
particles, initially with momenta ${\bf p}_1$ and ${\bf p}_2$, begin to have
${\bf p}_1^{\prime }$ and ${\bf p}_2^{\prime }$ momenta, respectively. If we
multiply this number by the flux of particle existing within this volume
\begin{equation}
\label{(47)}F_1\left( {\bf x},{\bf p}_1;t\right) \left| \frac{{\bf p}_1}{m_1}%
-\frac{{\bf p}_2}{m_2}\right| d^3p_1
\end{equation}
and by the number of {\it type2} particles which can bring forth such a
collision
\begin{equation}
\label{(48)}F_2\left( {\bf x},{\bf p}_2;t\right) d^3xd^3p_2,
\end{equation}
we obtain%
$$
D_C^{\left( -\right) }F_1\left( {\bf x},{\bf p}_1;t\right) =
$$
\begin{equation}
\label{(49)}=\int_{{\bf p}_1^{\prime }}\int_{{\bf p}_2^{\prime }}\int_{{\bf p%
}_2}\left| \frac{{\bf p}_1}{m_1}-\frac{{\bf p}_2}{m_2}\right| F_1\left(
1\right) F_2\left( 2\right) \sigma \left( {\bf p}_1,{\bf p}_2\rightarrow
{\bf p}_1^{\prime },{\bf p}_2^{\prime }\right) d^3p_1^{\prime
}d^3p_2^{\prime }d^3p_2,
\end{equation}
where $F_1\left( 1\right) $ and $F_2\left( 2\right) $ represent $F_1\left(
{\bf x},{\bf p}_1;t\right) $ and $F_2\left( {\bf x},{\bf p}_2;t\right) $,
respectively. In the same manner, using the inverse scattering
arrangement\cite{28}, we get for $D_C^{\left( +\right) }F_1$:%
$$
D_C^{\left( +\right) }F_1\left( {\bf x},{\bf p}_1;t\right) =
$$
\begin{equation}
\label{(50)}=\int_{{\bf p}_1^{\prime }}\int_{{\bf p}_2^{\prime }}\int_{{\bf p%
}_2}\left| \frac{{\bf p}_1^{\prime }}{m_1}-\frac{{\bf p}_2^{\prime }}{m_2}%
\right| F_1^1\left( 1^{\prime }\right) F_2^1\left( 2^{\prime }\right) \sigma
\left( {\bf p}_1^{\prime },{\bf p}_2^{\prime }\rightarrow {\bf p}_1,{\bf p}%
_2\right) d^3p_1^{\prime }d^3p_2^{\prime }d^3p_2,
\end{equation}
where $F_1^1\left( 1^{\prime }\right) $ and $F_2^1\left( 2^{\prime }\right) $
represent $F_1^1\left( {\bf x},{\bf p}_1^{\prime };t\right) $ and $%
F_2^1\left( {\bf x},{\bf p}_2^{\prime };t\right) $, respectively.

Using the fact that
\begin{equation}
\label{(51)}\sigma \left( {\bf p}_1,{\bf p}_2\rightarrow {\bf p}_1^{\prime },%
{\bf p}_2^{\prime }\right) =\sigma \left( {\bf p}_1^{\prime },{\bf p}%
_2^{\prime }\rightarrow {\bf p}_1,{\bf p}_2\right)
\end{equation}
and that, for elastic collisions,
\begin{equation}
\label{(52)}\left| \frac{{\bf p}_1}{m_1}-\frac{{\bf p}_2}{m_2}\right|
=\left| \frac{{\bf p}_1^{\prime }}{m_1}-\frac{{\bf p}_2^{\prime }}{m_2}%
\right| =\vartheta ,
\end{equation}
we finally have the factor%
$$
D_CF_1\left( 1\right) =\int_{{\bf p}_1^{\prime }}\int_{{\bf p}_2^{\prime
}}\int_{{\bf p}_2}\left| \frac{{\bf p}_1}{m_1}-\frac{{\bf p}_2}{m_2}\right|
\left( F_1\left( 1\right) F_2\left( 2\right) -F_1^1\left( 1^{\prime }\right)
F_2^1\left( 2^{\prime }\right) \right) \cdot
$$
\begin{equation}
\label{(53)}\cdot \sigma \left( {\bf p}_1^{\prime },{\bf p}_2^{\prime
}\rightarrow {\bf p}_1,{\bf p}_2\right) d^3p_1^{\prime }d^3p_2^{\prime
}d^3p_2.
\end{equation}

Taking this result to equation (\ref{(44)}), we have%
$$
\frac{\partial F_1^1}{\partial t}+\frac{{\bf p}_1}{m_1}\cdot \nabla _xF_1^1+
\frac{d{\bf p}_1}{dt}\cdot \nabla _pF_1^1=
$$
\begin{equation}
\label{(54)}=\int_{{\bf p}_1^{\prime }}\int_{{\bf p}_2^{\prime }}\int_{{\bf p%
}_2}\left( F_1\left( 1\right) F_2\left( 2\right) -F_1^1\left( 1^{\prime
}\right) F_2^1\left( 2^{\prime }\right) \right) \vartheta \sigma
d^3p_1^{\prime }d^3p_2^{\prime }d^3p_2
\end{equation}
and, after applying the Wigner-Moyal Infinitesimal Transformation, we obtain%
$$
\left\{ \frac{\hbar ^2}{2m_1}\left[ \nabla _{y_1}^2-\nabla _{y_1^{\prime
}}^2\right] -\left[ V\left( y_1\right) -V\left( y_1^{\prime }\right) \right]
\right\} \rho \left( 1\right) +i\hbar \frac \partial {\partial t}\rho \left(
1\right) =
$$
\begin{equation}
\label{(55)}=\int_{{\bf p}_1}\int_{{\bf p}_1^{\prime }}\int_{{\bf p}_2}\int_{%
{\bf p}_2^{\prime }}\left( F_1\left( 1\right) F_2\left( 2\right)
-F_1^1\left( 1^{\prime }\right) F_2^1\left( 2^{\prime }\right) \right)
\vartheta \sigma \left( {\bf p}_1^{\prime },{\bf p}_2^{\prime }\rightarrow
{\bf p}_1,{\bf p}_2\right) \cdot
\end{equation}
$$
\cdot \exp \left[ \frac i\hbar {\bf p}_1\cdot \left( {\bf y}_1-{\bf y}%
_1^{\prime }\right) \right] d^3p_1d^3p_1^{\prime }d^3p_2d^3p_2^{\prime },
$$
where, as usual, we did%
$$
{\bf y}_1={\bf x}_1+\frac{\delta {\bf x}_1}2\quad ;\quad {\bf y}_1^{\prime }=%
{\bf x}_1-\frac{\delta {\bf x}_1}2
$$
and $\rho \left( 1\right) =\rho \left( {\bf y}_1^{\prime },{\bf y}_1\right) $
refers, naturally, to {\it system1}.

It is obvious that we can invert the problem and interpret {\it system1} as
the observer and {\it system2} as the observed. In this case, we would have
for {\it system2} an equation similar to (\ref{(55)})%
$$
\left\{ \frac{\hbar ^2}{2m_2}\left[ \nabla _{y_2}^2-\nabla _{y_2^{\prime
}}^2\right] -\left[ V^{\prime }\left( y_2\right) -V^{\prime }\left(
y_2^{\prime }\right) \right] \right\} \rho \left( 2\right) +i\hbar \frac
\partial {\partial t}\rho \left( 2\right) =
$$
\begin{equation}
\label{(56)}=\int_{{\bf p}_2}\int_{{\bf p}_2^{\prime }}\int_{{\bf p}_1}\int_{%
{\bf p}_1^{\prime }}\left( F_1\left( 1\right) F_2\left( 2\right)
-F_1^1\left( 1^{\prime }\right) F_2^1\left( 2^{\prime }\right) \right)
\vartheta \sigma \left( {\bf p}_1^{\prime },{\bf p}_2^{\prime }\rightarrow
{\bf p}_1,{\bf p}_2\right) \cdot
\end{equation}
$$
\cdot \exp \left[ \frac i\hbar {\bf p}_2\cdot \left( {\bf y}_2-{\bf y}%
_2^{\prime }\right) \right] d^3p_2d^3p_2^{\prime }d^3p_1d^3p_1^{\prime },
$$
where $\rho \left( 2\right) =\rho \left( {\bf y}_2^{\prime },{\bf y}%
_2\right) $ refers to {\it system2} subject to a potential $V^{\prime
}\left( {\bf x}\right) $, possibly distinct from $V\left( {\bf x}\right) $
(we are supposing a purely contact based interaction between the different
particles and that there are no multiple collisions between the particles).

This ''symmetry'' between observer and observable is extremely important for
the special theory of relativity's point of view. It is also important to
observe that it is no longer possible to obtain an equation such as
Schroedinger's Second Equation from (\ref{(55)}) or (\ref{(56)}) (of course,
for a weak interaction between the two systems, such an equation can be
approximated). The ''wave'' properties associated to these particles should,
depending on the intensity of the interaction between both system,
disappear. We here note that this property, as all of this problem's
treatment, is quite distinct from the state vector's collapse.

We will discuss the epistemological implications of these and other
properties of equations (\ref{(55)}) and (\ref{(56)}) in the third paper of
this series.

\section{Conclusion}

{}From no more than three axioms and within a classical and coherent with
Realism view of nature, it was possible to derive all of quantum mechanical
formalism. It was also possible to derive the important issue of
Heisemberg's dispersion relations, one of the most fundamental result of the
Copenhagen Interpretation\cite{12}.

In the second paper of this series, we will undertake relativistic treatment.

In the final paper, we will discuss the epistemological implications of the
results that we have obtained.

\appendix{}

\section{Mathematical Properties of the Transformation}

The Wigner-Moyal Infinitesimal Transformation is defined as the following
Fou\-ri\-er Transform
\begin{equation}
\label{(A57)}\rho \left( x+\frac{\delta x}2,x-\frac{\delta x}2;t\right)
=\int F\left( x,p;t\right) \exp \left( i\frac{p\delta x}\hbar \right) dp.
\end{equation}
One could think, formally and at first sight, that the function $F\left(
x,p;t\right) $ could be obtained through the inverse Fourier Transform
\begin{equation}
\label{(A58)}F\left( x,p;t\right) =\int \rho \left( x+\frac{\delta x}2,x-
\frac{\delta x}2;t\right) \exp \left( -i\frac{p\delta x}\hbar \right)
d\left( \delta x\right) .
\end{equation}

Yet it is known\cite{21,22,29} that the function defined in (\ref{(A58)}) is
not positive-defined, and thus, the argument runs, it can not be considered
a true probability density.

Nevertheless, we should stress that it was necessary to consider $\delta x$
as being infinitesimal, in order for the density function to satisfy
Schroedinger's First Equation. In this manner, the transformation (\ref
{(A58)}) can not be performed. This means that, even possessing the solution
to the quantum problem, given by the density function, we can not obtain the
probability density $F\left( x,p;t\right) $ in phase space. On the other
hand, it is interesting to note that, if we have the solution in phase space
(usually called classical), we can obtain the density function (usually
called quantum) through the application of (\ref{(A57)}).

Even in view of the impossibility of the inverse transformation (\ref{(A58)}%
), we can easily demonstrate that
\begin{equation}
\label{(A59)}F_p\left( p;t\right) =\int F\left( x,p;t\right) dx\geq 0
\end{equation}
and, therefore, $F_p\left( p;t\right) $ serves as a probability density in
momentum space (however we must stress that this is presupposed in the
present formalism). In the same manner, it can be shown that
\begin{equation}
\label{(A60)}F_x\left( x;t\right) =\int F\left( x,p;t\right) dp\geq 0
\end{equation}
and, therefore, $F_x\left( x;t\right) $ serve as a probability density in
real space.

\subsection{Correspondence}

Let us suppose that we have a dispersion free {\it ensemble} of a single
particle system. In this case, the joint probability function is given as
\begin{equation}
\label{(A61)}F\left( x,p;t\right) =\delta \left( x-x_0\left( t\right)
\right) \delta \left( p-p_0\left( t\right) \right) ,
\end{equation}
meaning that, given the same initial conditions, the trajectory in phase
space followed by the particle will always be the same, given by $x_0\left(
t\right) $ and $p_0\left( t\right) $. The density function for this problem
is, using (\ref{(A57)})
\begin{equation}
\label{(A62)}\rho \left( x+\frac{\delta x}2,x-\frac{\delta x}2;t\right)
=\delta \left( x-x_0\left( t\right) \right) \exp \left( \frac i\hbar
p_0\left( t\right) \Delta x\right) ,
\end{equation}
where we use $\Delta x$ for infinitesimal dislocations so as to avoid
confusion with Dirac's delta distributions. Note that we can not write the
density function as the product of amplitudes. This is not unexpected, as
this function was derived from a probability density which did not satisfy
dispersion relations. Substituting this expression in equation (\ref{(6)}),
we obtain%
$$
i\hbar \left[ \frac{p_0\left( t\right) }m\frac \partial {\partial x}-\frac
\partial {\partial t}\right] \delta \left( x-x_0\left( t\right) \right) +
$$
\begin{equation}
\label{(A63)}+\Delta x\left[ \frac{dp_0\left( t\right) }{dt}+\left( \frac{%
\partial V}{\partial x}\right) _{x=x_0}\right] \delta \left( x-x_0\left(
t\right) \right) =0,
\end{equation}
which is a decomposition formally similar to that done in equation (\ref
{(14)}), in the appropriate limit, with the first term representing the
Continuity Equation and the last one, Schroedinger's Second Equation.

Noting that the real and complex parts should be equal to zero separately,
we obtain
\begin{equation}
\label{(A64)}\frac{dp_0\left( t\right) }{dt}=-\left( \frac{\partial V}{%
\partial x}\right) _{x=x_0},
\end{equation}
for the real term, and
\begin{equation}
\label{(A65)}\frac{p_0\left( t\right) }m=\left( \frac{dx}{dt}\right)
_{x=x_0},
\end{equation}
for the complex one. These are nothing more than Newton's equations
satisfied by each one of the {\it ensemble's} components.

It must be noted that the result obtained above {\it does not depend on
Plank's constant}. It is well stated in the existing literature that the
''classical limit'' is not always obtained when we make Plank's constant
tend to zero\cite{30}. Also note that, in the perspective of the present
work, the limit $\hbar \rightarrow 0$ would not make sense due to the
Wigner-Moyal Infinitesimal Transformation.

\section{Operator Construction}

One of the problems found within the usual formulation of quantum mechanics
concerns the construction of operators which represent a certain function in
the phase space\cite{19}. This problem is basically caused by the fact that
we usually work with operators which act upon the probability amplitude and
do not commute with each other. We have seen, in the third section of this
paper, that we can define position and momentum operators which act upon the
density function and commute with each other. Therefore we hope to be
capable of, given a function $f\left( {\bf x},{\bf p};t\right) $,
constructing an operator to represent it when acting upon the density
function. In fact, let $f\left( {\bf x},{\bf p};t\right) $ be a function
whose mean value we desire to calculate. In this case we have%
$$
\overline{f\left( {\bf x},{\bf p};t\right) }=\int \int f\left( {\bf x},{\bf p%
};t\right) F\left( {\bf x},{\bf p};t\right) d^3xd^3p=
$$
$$
=\lim _{\delta x\rightarrow 0}\int \int f\left( {\bf x},{\bf p};t\right)
F\left( {\bf x},{\bf p};t\right) \exp \left( i\frac{{\bf p}\cdot \delta {\bf %
x}}\hbar \right) d^3xd^3p=
$$
\begin{equation}
\label{(B66)}=\lim _{\delta x\rightarrow 0}\int \int O_p\left( {\bf x}%
,\delta {\bf x};t\right) \rho \left( x+\frac{\delta x}2,x-\frac{\delta x}%
2;t\right) d^3x,
\end{equation}
and can say that $O_p\left( {\bf x},\delta {\bf x};t\right) $ is the
operator associated to $f\left( {\bf x},{\bf p};t\right) $, with the process
of limit understood.

It is thus easy to demonstrate as an example that, for angular momentum,
\begin{equation}
\label{(B67)}{\bf L=x}\times {\bf p}
\end{equation}
we get for the components
\begin{equation}
\label{(B68)}\stackrel{\wedge }{L}_i^{\prime }=-i\hbar \epsilon
_{ijk}x_j\frac \partial {\partial \left( \delta x_k\right) },
\end{equation}
where we place an apostrophe on $L_i^{\prime }$ to mark that this operator
acts upon a density function and not upon the probability amplitude. We will
have, in general, the following correspondence rule
\begin{equation}
\label{(B69)}O_p\left( g\left( {\bf x},{\bf p}\right) \right) =g\left( {\bf x%
},-i\hbar \frac \partial {\partial \left( \delta {\bf x}\right) }\right) .
\end{equation}

The greatest difficult associated to the methods of operator construction in
usual quantum mechanics refers to a certain ambiguity which they present\cite{%
19} or their incompatibility to the one to one correspondence between
operators and observables postulated in quantum mechanics. Among the methods
proposed in the literature, we can cite: von Neumann's rules, Weyl's rules,
Revier's rules, etc. The ambiguity problem has, nevertheless, prevailed. In
fact, for the function $p^2x^2$, von Neumann's rules give, for example:
\begin{equation}
\label{(B70)}O\left( p^2x^2\right) =\stackrel{\wedge }{x}^2\stackrel{\wedge
}{p}^2-2i\hbar \stackrel{\wedge }{x}\stackrel{\wedge }{p}-\frac 14\hbar
^2\quad ;\quad O\left( p^2x^2\right) =\stackrel{\wedge }{x}^2\stackrel{%
\wedge }{p}^2-2i\hbar \stackrel{\wedge }{x}\stackrel{\wedge }{p}-\hbar ^2,
\end{equation}
where $O\left( p^2x^2\right) $ represents the operator associated to the
calculation of the mean value of his argument. According to the present
theory we have, naturally,
\begin{equation}
\label{(B71)}O^{\prime }\left( p^2x^2\right) =-\hbar ^2x^2\frac{\partial ^2}{%
\partial \left( \delta x_k\right) ^2},
\end{equation}
which, expanding the density function according to (\ref{(11)}) and
performing the calculations, reduces to the following operator for the
probability amplitude
\begin{equation}
\label{(B72)}O\left( p^2x^2\right) =\stackrel{\wedge }{x}^2\stackrel{\wedge
}{p}^2-i\hbar \stackrel{\wedge }{x}\stackrel{\wedge }{p},
\end{equation}
which does not present any ambiguity.

\section{Three-dimensional Formalism}

For a problem involving {\it N} particles the state of the ensemble is
represented by the function $F\left( {\bf x}_1,{\bf p}_1,..,{\bf x}_N,{\bf p}%
_N;t\right) $. For this function we will have
\begin{equation}
\label{(C73)}\frac{dF}{dt}=\frac{\partial F}{\partial t}+\sum_i\frac{d{\bf x}%
_i}{dt}\cdot \nabla _{x_i}F+\sum_i\frac{d{\bf p}_i}{dt}\cdot \nabla
_{p_i}F=0.
\end{equation}

We can use
\begin{equation}
\label{(C74)}\frac{d{\bf x}_i}{dt}=\frac{{\bf p}_i}m\quad ;\quad \frac{d{\bf %
p}_k}{dt}={\bf f}_k^i+{\bf f}_k^e,
\end{equation}
where ${\bf f}_k^i$ are the internal forces represented by
\begin{equation}
\label{(C75)}{\bf f}_k^i=\sum_{l\neq k}{\bf f}_{l\rightarrow k}^i\left( {\bf %
x}_{kl}\right) \quad ;\quad {\bf x}_{kl}={\bf x}_k-{\bf x}_l,
\end{equation}
with ${\bf f}_{l\rightarrow k}^i$ representing the internal force exercised
by particle {\it l} upon particle {\it k}, depending only on the relative
position, and ${\bf f}_k^e$ are the external forces acting on particle {\it k%
}.

Using the potentials
\begin{equation}
\label{(C76)}{\bf f}_k^i=-\nabla _{x_k}V^i\quad ;\quad V^i=\frac
12\sum_{l\neq k}V_{kl}^i\left( {\bf x}_{kl}\right)
\end{equation}
and
\begin{equation}
\label{(C77)}{\bf f}_k^e=-\nabla _{x_k}V^e,
\end{equation}
jointly with the Wigner-Moyal Infinitesimal Transformation%
$$
\rho \left( {\bf x}_1+\frac{\delta {\bf x}_1}2,{\bf x}_1-\frac{\delta {\bf x}%
_1}2,..,{\bf x}_N+\frac{\delta {\bf x}_N}2,{\bf x}_N-\frac{\delta {\bf x}_N}%
2;t\right) =
$$
\begin{equation}
\label{(C78)}=\int ..\int F\left( {\bf x}_1,{\bf p}_1,..,{\bf x}_N,{\bf p}%
_N;t\right) \exp \left[ \frac i\hbar \left( {\bf p}_1\cdot \delta {\bf x}%
_1+..+{\bf p}_N\cdot \delta {\bf x}_N\right) \right] d^3p_1..d^3p_N,
\end{equation}
we reach the equation%
$$
\left\{ \frac{\hbar ^2}{2m}\left[ \nabla _{y_k}^2-\nabla _{y_k^{\prime
}}^2\right] -\sum_l\left[ V^i\left( {\bf y}_{kl}\right) -V^i\left( {\bf y}%
_{kl}^{\prime }\right) \right] \right\} \rho -
$$
\begin{equation}
\label{(C79)}-\sum_l\left[ V^e\left( {\bf y}_{kl}\right) -V^e\left( {\bf y}%
_{kl}^{\prime }\right) \right] \rho =-i\hbar \frac \partial {\partial t}\rho
,
\end{equation}
where we have made the usual variable transformations. Equation (\ref{(C79)}%
) is the Schroedinger's First Equation for the density function for an {\it %
ensemble} built of {\it N} particle systems. Schroedinger's Second Equation
can be obtained from (\ref{(C79)}) with calculations similar to those
realized for the one dimensional problem.

\section{Density Matrix}

Until now we have only dealt with {\it ensembles} which can be represented
by {\it pure states}\cite{31}. In this appendix we will proceed to
generalize the theory for {\it mixed states}. We will also show that the
procedure usually followed in the literature involves an additional
assumption. Indeed, when developing the density matrix theory one usually
begins with the amplitudes which are solutions of the Second Schroedinger's
Equation $\left| \Psi \right\rangle $ and then define the density function
as the product
\begin{equation}
\label{(D0)}\left| \Psi \left( y\right) \right\rangle \left\langle \Psi
\left( y^{\prime }\right) \right|
\end{equation}
where $y$ and $y^{\prime }$ run independently. We will suppose throughout
this appendix that it is always possible to write the density function as
the product (\ref{(11)}) and will show that (\ref{(D0)}) is related to this
additional assumption.

It was seen for pure states (see appendix B) that, when the decomposition
given by expression (\ref{(11)}) is possible, we can always write an
operator acting upon the density function $\stackrel{\wedge }{Q}^{\prime }$%
as another one acting upon the probability amplitudes $\stackrel{\wedge }{Q}$%
, mathematically:%
$$
\left\langle Q\right\rangle =\lim _{\delta x\rightarrow 0}\int \stackrel{%
\wedge }{Q}^{\prime }\rho \left( x-\frac{\delta x}2,x+\frac{\delta x}%
2;t\right) dx=
$$
\begin{equation}
\label{(D1)}=\int \Psi ^{*}\left( x;t\right) \stackrel{\wedge }{Q}\Psi
\left( x;t\right) dx,
\end{equation}
where
\begin{equation}
\label{(D2)}\rho \left( x-\frac{\delta x}2,x+\frac{\delta x}2;t\right) =\Psi
^{*}\left( x-\frac{\delta x}2;t\right) \Psi \left( x+\frac{\delta x}%
2;t\right)
\end{equation}
is the density function representing the pure state defined by $\Psi \left(
x;t\right) $.

The generalization for mixed states can be done writing the density function
as
\begin{equation}
\label{(D3)}\rho \left( x-\frac{\delta x}2,x+\frac{\delta x}2;t\right)
=\sum_nW_n\Psi _n^{*}\left( x-\frac{\delta x}2;t\right) \Psi _n\left( x+
\frac{\delta x}2;t\right) ,
\end{equation}
where the{\it \ }$W_n$ are the statistical weights. It is easy to see that
$$
\left\langle Q\right\rangle =\lim _{\delta x\rightarrow 0}\int \stackrel{%
\wedge }{Q}^{\prime }\rho \left( x-\frac{\delta x}2,x+\frac{\delta x}%
2;t\right) dx=
$$
\begin{equation}
\label{(D4)}=\sum_nW_n\int \Psi _n^{*}\left( x;t\right) \stackrel{\wedge }{Q}%
\Psi _n\left( x;t\right) dx.
\end{equation}

We can now choose a convenient representation for our amplitudes using some
set of orthonormal basis states $\left\{ \phi _i\right\} \quad i=1,2,..$ for
which
\begin{equation}
\label{(D5)}\Psi _n\left( x;t\right) =\left| \Psi _n\right\rangle
=\sum_ma_m^{(n)}\left| \phi _m\right\rangle ,
\end{equation}
where we used Dirac's representation of Bras and Kets to simplify the
calculations. With expression (\ref{(D5)}), the mean value of the operator $%
Q $ can be written as
\begin{equation}
\label{(D6)}\left\langle Q\right\rangle =\sum_{n,m^{\prime
},m}W_na_{m^{\prime }}^{(n)*}a_m^{(n)}\left\langle \phi _m^{\prime }\right|
Q\left| \phi _m\right\rangle .
\end{equation}

Now we can {\it define} our {\it density matrix}, in the representation
given by $\left\{ \phi _i\right\} $, as the matrix which elements are given
by
\begin{equation}
\label{(D7)}\stackrel{\longleftrightarrow }{\rho }_{m^{\prime
},m}=\sum_nW_na_{m^{\prime }}^{(n)*}a_m^{(n)}.
\end{equation}
Since, by means of (\ref{(D5)}) and orthonormality, we have
\begin{equation}
\label{(D8)}a_m^{(n)}=\left\langle \phi _m\right| \left. \Psi
_n\right\rangle \quad and\quad a_{m^{\prime }}^{(n)*}=\left\langle \Psi
_n\right| \left. \phi _{m^{\prime }}\right\rangle ,
\end{equation}
it is possible to write
\begin{equation}
\label{(D9)}\stackrel{\longleftrightarrow }{\rho }_{m,m^{\prime
}}=\sum_nW_n\left\langle \phi _m\right| \left. \Psi _n\right\rangle
\left\langle \Psi _n\right| \left. \phi _{m^{\prime }}\right\rangle ,
\end{equation}
by which the density function follows
\begin{equation}
\label{(D10)}\rho \left( y^{\prime },y\right) =\sum_nW_n\left| \Psi _n\left(
y^{\prime }\right) \right\rangle \left\langle \Psi _n\left( y\right) \right|
,
\end{equation}
where $y$ and $y^{\prime }$ should be independent variables for the dot
product implied by (\ref{(D9)}) to be correct. Using results (\ref{(D6)})-(%
\ref{(D10)}) we can see that we have
\begin{equation}
\label{(D11)}\left\langle Q\right\rangle =Tr\left( \stackrel{%
\longleftrightarrow }{\rho }Q\right) .
\end{equation}

In the above derivation, we begin with the density function and an operator
acting upon it and then, supposing decomposition (\ref{(D2)}), turned into a
formalism with probability amplitudes and the related modified operator.
When making such calculations it was necessary to take the limit $\delta
x\rightarrow 0$. This limit implies changing from a commutative formalism
into a non--commutative one, as is stated by theorem (T4). Expression (\ref
{(D4)}) and (\ref{(D11)}) are equivalent only if it is possible to consider
{\it y} and {\it y'} as independent variables (not necessarily
infinitesimally separated). That such a supposition is made when we use
decomposition (\ref{(D2)}) will now be demonstrated.

When decomposition (\ref{(D2)}) is assumed, it is easy to show that the
density function equation can be written as%
$$
\left\{ \frac 1\Psi \left[ H\Psi -i\hbar \frac{\partial \Psi }{\partial t}%
\right] \right\} \left( x+\frac{\delta x}2\right) -\left\{ \frac 1\Psi
\left[ H\Psi -i\hbar \frac{\partial \Psi }{\partial t}\right] \right\}
\left( x-\frac{\delta x}2\right) =
$$
\begin{equation}
\label{(D12)}=\delta x\frac \partial {\partial x}\left\{ \frac 1\Psi \left[
H\Psi -i\hbar \frac{\partial \Psi }{\partial t}\right] \right\} \left(
x\right) =0,
\end{equation}
where $H$ is the Hamiltonian defined in (\ref{(27)}). But if we say that
\begin{equation}
\label{(D13)}H\Psi \left( x\right) -i\hbar \frac{\partial \Psi \left(
x\right) }{\partial t}=0\quad ,\forall x\in R,
\end{equation}
then equation (\ref{(D12)}) is satisfied independently of the infinitesimal
parameter. This ends the demonstration.

So we conclude that, only when it is possible to write the density function
as in (\ref{(D2)}), the mean values of the operators $Q$ can be calculated
using formulae (\ref{(D5)})-(\ref{(D11)}), as is usually done in the
literature.

\newpage

\unitlength=1.00mm
\special{em:linewidth 1pt}
\linethickness{1pt}

\begin{figure}
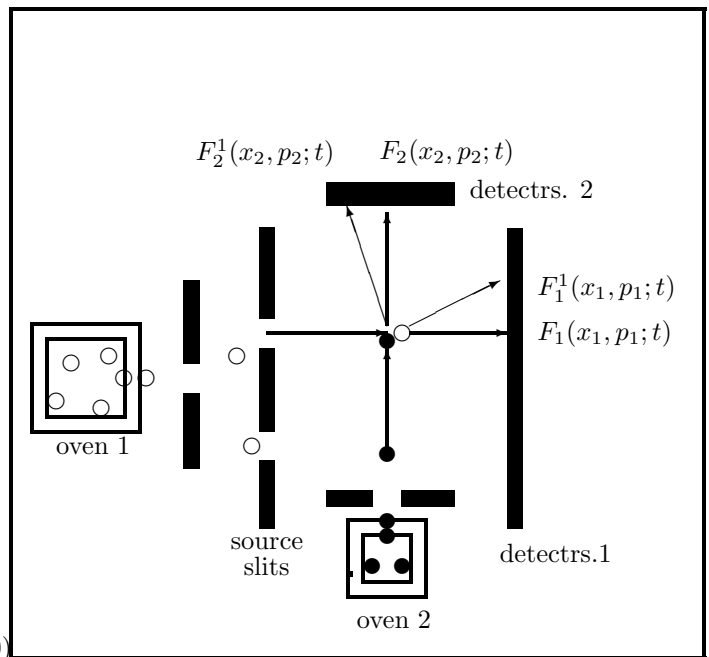
(92.00,86.00)
\put(5.00,32.00){\framebox(10.00,10.00)[cc]{}}
\put(3.00,30.00){\framebox(14.00,14.00)[cc]{}}
\put(23.00,39.00){\rule{2.00\unitlength}{11.00\unitlength}}
\put(23.00,25.00){\rule{2.00\unitlength}{10.00\unitlength}}
\put(33.00,45.00){\rule{2.00\unitlength}{12.00\unitlength}}
\put(33.00,30.00){\rule{2.00\unitlength}{11.00\unitlength}}
\put(33.00,17.00){\rule{2.00\unitlength}{9.00\unitlength}}
\put(45.00,11.00){\framebox(0.00,0.00)[cc]{}}
\put(45.00,8.00){\framebox(10.00,10.00)[cc]{}}
\put(47.00,10.00){\framebox(6.00,6.00)[cc]{}}
\put(42.00,20.00){\rule{6.00\unitlength}{2.00\unitlength}}
\put(52.00,20.00){\rule{7.00\unitlength}{2.00\unitlength}}
\put(66.00,17.00){\rule{2.00\unitlength}{40.00\unitlength}}
\put(8.00,39.00){\circle{2.00}}
\put(12.00,33.00){\circle{2.00}}
\put(13.00,40.00){\circle{2.00}}
\put(6.00,34.00){\circle{2.00}}
\put(15.00,37.00){\circle{2.00}}
\put(18.00,37.00){\circle{2.00}}
\put(30.00,40.00){\circle{2.00}}
\put(50.00,27.00){\circle*{2.00}}
\put(50.00,16.00){\circle*{2.00}}
\put(50.00,18.00){\circle*{2.00}}
\put(52.00,12.00){\circle*{2.00}}
\put(48.00,12.00){\circle*{2.00}}
\put(52.00,43.00){\circle{2.00}}
\put(50.00,42.00){\circle*{2.00}}
\put(34.00,43.00){\vector(1,0){16.00}}
\put(50.00,28.00){\vector(0,1){13.00}}
\put(42.00,60.00){\rule{17.00\unitlength}{3.00\unitlength}}
\put(53.00,44.00){\vector(2,1){12.00}}
\put(50.00,44.00){\vector(-1,3){5.33}}
\put(32.00,28.00){\circle{2.00}}
\put(6.00,27.00){\makebox(0,0)[lb]{oven 1}}
\put(46.00,6.00){\makebox(0,0)[lt]{oven 2}}
\put(65.00,15.00){\makebox(0,0)[lt]{detectrs.1}}
\put(61.00,61.00){\makebox(0,0)[lb]{detectrs. 2}}
\put(49.00,67.00){\makebox(0,0)[lc]{$F_2(x_2,p_2;t)$}}
\put(43.00,67.00){\makebox(0,0)[rc]{$F_{2}^{1}(x_2,p_2;t)$}}
\put(34.00,15.00){\makebox(0,0)[cc]{source}}
\put(34.00,12.00){\makebox(0,0)[cc]{slits}}
\put(50.00,44.00){\vector(0,1){15.00}}
\put(53.00,43.00){\vector(1,0){13.00}}
\put(70.00,49.00){\makebox(0,0)[lc]{$F_{1}^{1}(x_1,p_1;t)$}}
\put(70.00,43.00){\makebox(0,0)[lc]{$F_1(x_1,p_1;t)$}}
\put(0.00,0.00){\framebox(92.00,86.00)[cc]{}}

\caption{Experimental environment for the introduction of the
observer into quantum mechanical formalism.}
\end{figure}

\end{document}